\documentclass{article}

\usepackage{arxiv}
\usepackage[utf8]{inputenc} 
\usepackage[T1]{fontenc}    
\PassOptionsToPackage{hyphens}{url}
\usepackage{hyperref}       
\usepackage{url}            
\usepackage{booktabs}       
\usepackage{amsfonts}       
\usepackage{nicefrac}       
\usepackage{amsmath,amssymb}
\usepackage{cleveref}       
\usepackage[square,numbers]{natbib}
\usepackage{doi}
\usepackage{graphicx}
\usepackage[subrefformat=parens]{subcaption}
\usepackage{threeparttable}

\title{ABCI 3.0: Evolution of the leading AI infrastructure in Japan}

\date{November 14, 2024}

\usepackage{authblk}

\if0
\usepackage[backend=biber, style=numeric, sorting=none, natbib=true]{biblatex}
\DeclareSourcemap{
	\maps[datatype=bibtex]{
		\map[overwrite=false]{
			\step[fieldsource=note]
			\step[fieldset=addendum, origfieldval, final]
			\step[fieldset=note, null]
		}
	}
}
\renewbibmacro{in:}{%
\ifentrytype{inproceedings}{%
  \setunit{}
  \addperiod\addspace In \textit{Proc.\ of the}}%
{\printtext{\bibstring{in}\intitlepunct}}
}
\DeclareFieldFormat{volume}{#1}
\fi

\author[1]{Ryousei Takano\thanks{\texttt{takano-ryousei@aist.go.jp}}}
\author[1]{Shinichiro Takizawa}
\author[1]{Yusuke Tanimura}
\author[1]{Hidemoto Nakada}
\author[2]{Hirotaka Ogawa}
\affil[1]{National Institute of Advanced Industrial Science and Technology (AIST), Japan.}
\affil[2]{AIST Solutions Co., Japan.}

\renewcommand{\headeright}{}
\renewcommand{\undertitle}{}
\renewcommand{\shorttitle}{ABCI 3.0}

\hypersetup{
pdftitle={ABCI 3.0: Evolution of the leading AI infrastructure in Japan},
pdfsubject={cs.DC, cs.AI},
pdfauthor={Ryousei Takano, et. al.},
pdfkeywords={Generative AI, AI Supercomputer, HPC},
}

\begin{document}
\maketitle

\begin{abstract}
ABCI 3.0 is the latest version of the ABCI, a large-scale open AI infrastructure that AIST has been operating since August 2018 and will be fully operational in January 2025.
ABCI 3.0 consists of computing servers equipped with 6128 of the NVIDIA H200 GPUs and an all-flash storage system. Its peak performance is 6.22 exaflops in half precision and 3.0 exaflops in single precision, which is 7 to 13 times faster than the previous system, ABCI 2.0. It also more than doubles both storage capacity and theoretical read/write performance.
ABCI 3.0 is expected to accelerate research and development, evaluation, and workforce development of cutting-edge AI technologies, with a particular focus on generative AI.
\end{abstract}


\newcommand{\ctext}[1]{\raise0.2ex\hbox{\textcircled{\scriptsize{#1}}}}


\section{Introduction}
AI Bridging Cloud Infrastructure (ABCI) is the world's first large-scale open AI infrastructure, designed and developed by AIST with the aim of accelerating the development of AI technologies in Japan~\cite{ABCI,Ogawa2018.SWoPP,Takizawa2021.SWoPP}.
It has been installed in the AI Data Center located at AIST's Kashiwa Center, and began operating in August 2018.
To date, many organizations have achieved remarkable results by using ABCI, including the successful construction of Japanese large language models~(LLMs) such as PLaMo~\cite{PLaMo}, Swallow LLM~\cite{Swallow}, and so on.
On the other hand, as the demand for generative AI in industry, academia, and government in Japan is growing rapidly, ABCI cannot provide enough resources to meet such user demands in a timely manner.
In addition, the development of generative AI is still at an early stage and currently focuses mainly on natural languages. In the near future, it is important to develop real world multimodal foundation models that are built using large amounts of image, audio and sensor data obtained from the real world, such as manufacturing, transportation, and so on.
In order to demonstrate such cutting-edge AI technologies, it is imperative to improve the computational capability of ABCI.
Since November 2024, AIST has been gradually starting to operate the latest system in the ABCI series, ABCI 3.0, while leveraging the existing technical assets of ABCI. ABCI 3.0 will be fully operational in January 2025.

This paper is organized as follows. Section~\ref{sec:hw} introduces the hardware configuration of ABCI 3.0. In Section~\ref{sec:sw}, we describe the initial software installed. The data center facility where we have upgraded for ABCI 3.0 is presented in Section~\ref{sec:facility}. Finally, Section~\ref{sec:future} briefly mentions the future perspective.

\begin{table}[tbh]
    \caption{ABCI series}
    \label{tbl:abci}
    \centering
    \begin{threeparttable}
    \begin{tabular}{p{2.5cm}|p{2.6cm}p{2.6cm}p{2.9cm}p{2.9cm}}
    \hline \hline
    & AAIC~\tnote{a} & ABCI 1.0 & ABCI 2.0~\tnote{b} & ABCI 3.0 \\
    \hline
    Start of operation & 2017 & 2018 & 2021 & 2024 \\
    Number of nodes & 50 & 1088 & 120 & 766 \\
    GPU & NVIDIA P100 & NVIDIA V100 & NVIDIA A100 & NVIDIA H200 \\
    Number of GPUs & 400 & 4352 & 960 & 6128 \\
    CPU & Intel Xeon E5 v4 & Intel Xeon Gold 6148 & Intel Xeon Platinum 8360Y & Intel Xeon Platinum 8558 \\
    FP64 (PFLOPS) & 2.2 & 37.2 & 19 & 415 \\
    FP32 (PFLOPS) & 4.4 & 74.4 & 150 & 3000 \\
    FP16 (PFLOPS) & 8.6 & 550 & 300 & 6220 \\
    \hline
    \end{tabular}
    \begin{tablenotes}
        \item[a]{AAIC is the prototype system of ABCI.}
        \item[b]{ABCI 2.0 is the extension based on ABCI 1.0, and this table only shows the specification of the extension part. We usually refer to the performance of ABCI 2.0 as the performance of ABCI 1.0 with the extension part.}
    \end{tablenotes}
    \end{threeparttable}
\end{table}

\section{ABCI 3.0 Hardware}\label{sec:hw}
\begin{figure}[t!]
    \centering
    \includegraphics[width=.6\columnwidth]{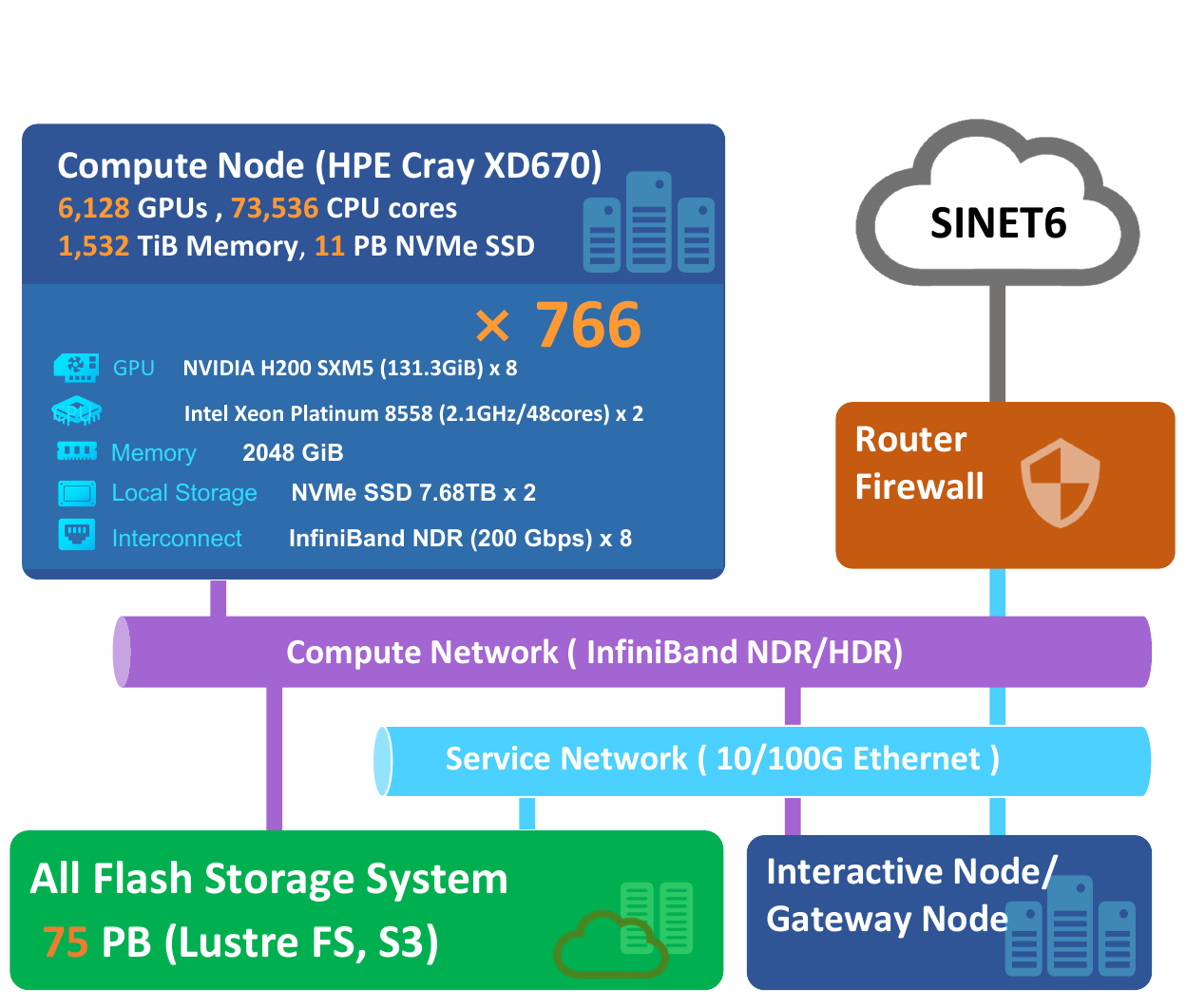}
    \caption{ABCI 3.0 hardware configuration}
    \label{fig:hw}
\end{figure}

Figure~\ref{fig:hw} shows the overview of ABCI 3.0 hardware configuration.
Table~\ref{tbl:abci} shows the specifications for the ABCI series, from AAIC, which is the prototype system of ABCI, to ABCI 3.0.
ABCI 3.0 consists of; 766 Compute Nodes (H) that form in total 6128 NVIDIA H200 GPU, shared file systems and cloud storage that provide in total 75PB capacity, InfiniBand NDR/HDR networks connecting the compute nodes and the storage systems, firewall equipments, and etc. ABCI 3.0 connects to the Internet through SINET6, the Science Information NETwork, at 400Gbps.
Its peak performance is 6.22 exaflops in half precision and 3.0 exaflops in single precision, which is 7 to 13 times faster than the previous system, ABCI 2.0. It also more than doubles both storage capacity and theoretical read/write performance.

The compute node consists of HPE Cray XD670, that is equipped with two Intel Xeon Platinum 8558 Processors (2.1GHz, 48core) and eight NVIDIA H200 SXM GPUs, 2048 GiB DDR5-5600 memory, two 7.68TB U.3 NVMe SSDs, eight InfiniBand NDR200 adapters, and an InfiniBand HDR adapter.

The interconnection network adopts a three-tier fat-tree topology, and all compute nodes are connected with full bisection bandwidth. Each compute node has eight InfiniBand NDR200 HBAs, and the total injection bandwidth is 12 times larger than that of ABCI 2.0. Compute nodes and the storage system are connected via InfiniBand HDR network.

ABCI 3.0 achieves a highly cost-effective all-flash storage system by adopting QLC~(Quad-Level Cell, 4-bit MLC) SSDs. The storage system provides Lustre parallel file system and AWS S3 compatible object storage.

ABCI 3.0 also has Interactive Nodes and Gateway Nodes. Interactive nodes are servers to login and used to compile programs and submit jobs to compute nodes. Gateway nodes are NAT servers that are located between compute nodes and the Internet. Compute nodes are not directly connected to the Internet, their access to the Internet is relayed through gateway nodes.

\section{ABCI 3.0 Software}\label{sec:sw}

ABCI 3.0 provides a set of system software that makes maximum use of hardware resources and realizes advanced development of foundation models, and cloud operation.
Table~\ref{tbl:sw} shows the list of major software. Note that this is just a list at the initial installation time and will be updated in the future.
In order to facilitate easy migration from the previous system, the usage model is basically the same, including operating system, job scheduler, a temporal shared file system using local SSDs~(BeeOND), and container support. A notable additional feasure to the previous system is Open OnDemand~\cite{Hudak2018.OOD,Nakada2024.SIGHPC193} support, which provides users with a way to access computing resources via a web browser instead of SSH.

\begin{table}[tbh]
    \caption{ABCI 3.0 major software}
    \label{tbl:sw}
    \centering
    \begin{tabular}{llll}
    \hline \hline
    Name & Compute node & Interactive node & Web service node \\
    \hline
    Rocky Linux & $\checkmark$ & - & - \\
    RedHat Enterprise Linux & - & $\checkmark$ & $\checkmark$ \\
    HPE Performance Cluster Manager & $\checkmark$ & $\checkmark$ & $\checkmark$ \\
    Altair PBS Professional & $\checkmark$ & $\checkmark$ & - \\
    Open OnDemand & - & - & $\checkmark$ \\
    Singularity CE & $\checkmark$ & - & - \\
    \hline
    DDN EXAScaler (Lustre) & $\checkmark$ & $\checkmark$ & $\checkmark$ \\
    DDN S3 API & $\checkmark$ & $\checkmark$ & - \\
    BeeOND & $\checkmark$ & - & - \\
    \hline
    NVIDIA CUDA Toolkit & $\checkmark$ & $\checkmark$ & - \\
    NVIDIA HPC SDK & $\checkmark$ & $\checkmark$ & - \\
    Intel oneAPI & $\checkmark$ & $\checkmark$ & - \\
    \hline
    \end{tabular}
\end{table}

In order to achieve high resource utilization, ABCI 3.0 uses Altair PBS Professional job scheduler for resource management. Each compute node has eight GPUs, but some workloads may require fewer GPUs and/or CPU cores. ABCI 3.0 allocates the same compute node to multiple jobs without resource oversubscription, with the job scheduler dynamically allocating a partial resources using cgroups and disk quota. Table~\ref{tbl:rtype} summarises ABCI 3.0 resource types. It intends a single compute node (rt\_HF) can be divided into eight rt\_HG instances and two rt\_HC instances.

Unlike HPC, AI related software updates frequently, making it difficult for system operators to prepare all the necessary software in advance. Therefore, it is necessary for users to be able to freely customize their software environment. One such mechanism is environment modules, and the other is containers. We provide several environment modules to enable users easily use various kinds and versions of software libraries, including CUDA, cuDNN, NCCL, MPI, and Python. Users can build their development environments by loading environment modules and then installing necessary Python libraries such as PyTorch, DeepSpeed, Hugging Face Hub by running pip, Anaconda, or similar tools.
There are several container runtime systems that integrates well with HPC environments. For example, Singularity, Podman, and Enroot. ABCI 3.0 initially support Singularity Community Edition~(CE).
In addition, Spack software package manager helps users compile and install software that is not officially supported by ABCI.

\begin{table}[tbh]
    \caption{ABCI 3.0 Resource types}
    \label{tbl:rtype}
    \centering
    \begin{tabular}{p{2cm}|lrrrr}
    \hline \hline
    Resource type & Description & \#CPU core & \#GPU core & Memory (GiB) & Local storage (TB) \\
    \hline
    rt\_HF & Node-exclusive & 96 & 8 & 1728 & 14  \\
    rt\_HG & Node-sharing with 1 GPU & 8  & 1 & 144  & 1.4 \\
    rt\_HC & Node-sharing with CPU only & 16 & 0 & 288  & 1.4 \\
    \hline
    \end{tabular}
\end{table}

\section{AI Data Center Facility}\label{sec:facility}

\begin{figure}[t!]
    \centering
    \includegraphics[width=.5\columnwidth]{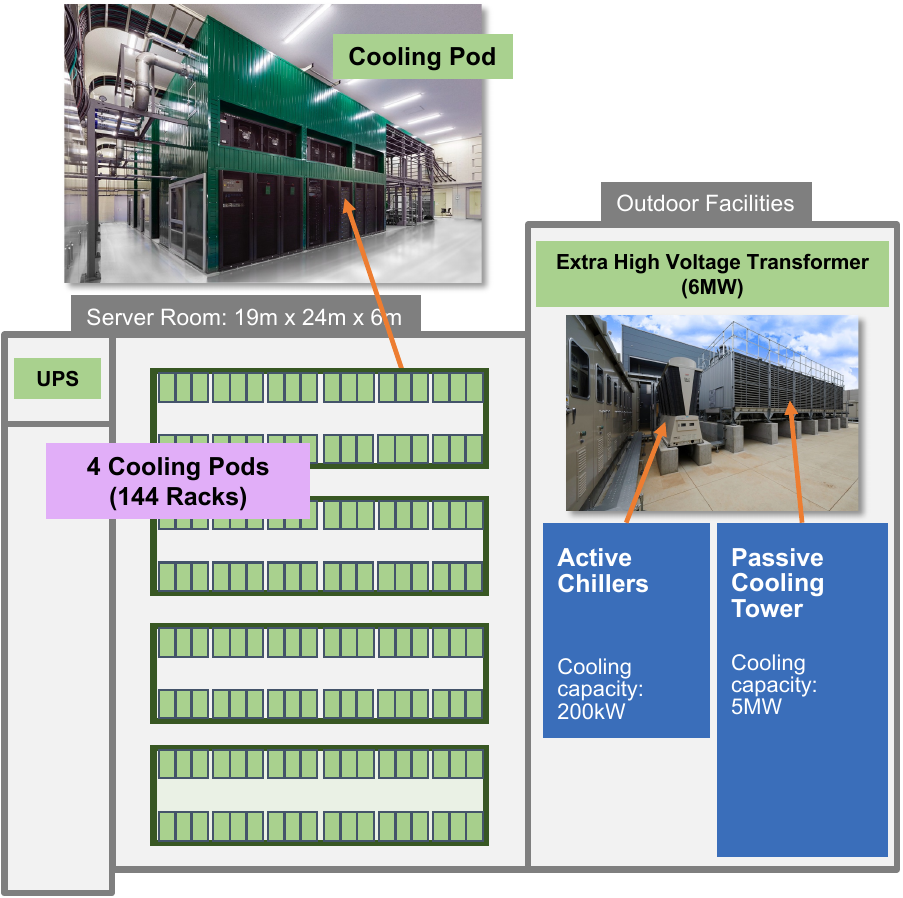}
    \caption{AI data center facility}
    \label{fig:aidc}
\end{figure}

Figure~\ref{fig:aidc} shows the overview of our AI data center facility that was built at the same time as ABCI 1.0. We have reported the detailed design of the original AI data center in \cite{Takano2018.SWoPP}.
For the installation of ABCI 3.0, we have upgraded the electrical and cooling capacities. Currently, we have 6MW of electrical capacity, 5.2MW of cooling capacity, and 144 racks of space.

\section{Future Perspective}\label{sec:future}
ABCI 1.0 and 2.0 are very successful systems for advancing AI R\&D in Japan. However, in the era of generative AI, AI industry is changing rapidly, and the situation surrounding ABCI is completely different from when we started the ABCI project. With government and other investment, the presence of domestic commercial cloud providers is growing. As a result of this situation, the role of ABCI must be shifted. ABCI 3.0 should focus on challenging research and development based on national demands, such as sovereign AI, and encouraging AI startups should also be an important mission. We provide large-scale computing resources needed for more challenging R\&D by relaxing the current usage limits, including the number of nodes per job, node-time product per job, maximum reserved duration, and etc. We also plan to conduct development acceleration programs and user events.
The latest information is available on the ABCI homepage~\cite{ABCI}.


\bibliographystyle{unsrtnat}
\bibliography{abci}

\end{document}